\RequirePackage{lineno}

\documentclass[twocolumn,aps,prc,showpacs,superscriptaddress,floatfix]{revtex4}

\usepackage{color}
\usepackage{graphicx}
\usepackage{multirow}
\usepackage{amsmath}
\usepackage{diagbox}

\begin{document}
\title{Solving Schrodinger equations using physically constrained neural network}
\author{Kai-Fang Pu}
\address{College of Science, Wuhan University of Science and Technology, Wuhan, Hubei 430065, China}
\author{Hanlin Li}
\email{lihl@wust.edu.cn}
\address{College of Science, Wuhan University of Science and Technology, Wuhan, Hubei 430065, China}
\author{Hong-Liang L\"u}
\address{HiSilicon Research Department, Huawei Technologies Co., Ltd., Shenzhen 518000, China}
\author{Long-Gang Pang}
\email{lgpang@ccnu.edu.cn}
\address{ Key Laboratory of Quark and Lepton Physics (MOE) and Institute of Particle Physics, Central China Normal University, Wuhan 430079, China}

\date{\today}

\begin{abstract}
  Deep neural network (DNN) and auto differentiation have been widely used in computational physics to solve variational problems. 
When DNN is used to represent the wave function to solve quantum many-body problems using variational optimization, 
various physical constraints have to be injected into the neural network by construction, to increase the data and learning efficiency.
We build the unitary constraint to the variational wave function using a monotonic neural network to represent the Cumulative Distribution Function (CDF)  $F(x) = \int_{-\infty}^{x} \psi^*\psi dx'$.
Using this constrained neural network to represent the variational wave function, we solve Schrodinger equations using auto-differentiation and stochastic gradient descent (SGD), by minimizing the violation of the trial wave function $\psi(x)$ to the Schrodinger equation. 
For several classical problems in quantum mechanics, we obtain their ground state wave function and energy with very low errors. The method developed in the present paper may pave a new way in solving nuclear many body problems in the future.

\par\textbf{Keywords: } Deep neural network; auto differentiation; variational problems; the Cumulative Distribution Function;  ground state wave function

\end{abstract}

\maketitle

\section{\label{sec:level1}I\lowercase{ntroduction}}

The universal approximation theorem of deep neural network (DNN) \cite{citeulike:3561150} makes it powerful in representing a variational function $y = f(x, \theta)$ with trainable parameters $\theta$.
In physics, this function can be used as solutions of many different partial differential equations (PDEs) $\hat{L} f = 0$, such as Maxwell equations for electromagnetic field, Navier-stokes equations for fluid dynamics, Schrodinger equations in quantum mechanics as well as the Einstein field equations for gravity.The traditional way to solve this problem is to use physical models. These models face great challenges in solving inverse problems with complex geometric regions and high-dimensional space. Unlike these models, the deep learning method developed in this paper provides a new direction to solve these problems.
As the parameters of DNN are initialized with random numbers, the variational function $f(x, \theta)$ violates PDEs and the residuals $\delta = |\hat{L} f|$ are usually the optimization objectives that can be minimized to desired precision. In this way, many physical problems \cite{RevModPhys.94.031003} are naturally mapped into optimization problems \cite{mackay2003information} that can be solved using the modern deep learning libraries. 

The main advantages of machine learning are that (1) it directly establishes the function mapping between input and output data, (2) ordinary differential equations (ODEs), PDEs can be transformed into variational problems that can be solved using optimization. It can be helpful in finding low-dimensional manifolds in high-dimensional space, which is crucial for the quantum many-body problem, which suffers from the curse of dimensionality. The associated disadvantage is that it is at an early stage of development and its applicability to computational physics has not been fully tested.

With strong information encapsulation capability, deep learning is proved to be a powerful tool in solving quantum many-body problems \cite{Mehta2018AHL,Nordhagen2022EfficientSO,Barrett:2013nh,torlai2018neural,PhysRevLett.127.022502}. The most typical application is to use DNN to represent the wave function of quantum many-body states for many-electron systems \cite{RevModPhys.91.045002}. In subsequent developments, ANN applications extended to prototypical spin lattice systems and  quantum systems in continuous space \cite{PhysRevLett.120.205302,Han2018SolvingMS,PhysRevD.105.014017}. Recently machine learning has been used to deal with ab-initio problems \cite{Choo2019FermionicNS,Scherbela2021SolvingTE,Yang2022DeepneuralnetworkSO}. The Feynman path integral \cite{RevModPhys.20.367} is another method for solve quantum state problems. Modern generative models can represent probability distribution with high computational efficiency. A Fourier-flow generative model has been proposed to simulate the Feynman propagator and generate paths for quantum systems \cite{Chen2022FourierflowMG}. And Ref \cite{Che2022EstimatingTE} propose a Feynman path generator that can estimate the Euclidean propagator and the ground state wave function with high accuracy.

PDEs usually have boundary and/or initial conditions. In an early paper, these initial and boundary conditions are built into the neural network through construction, the training objective is to minimize the residual $\delta$ alone. This method uses hard constraints such that $f(x, \theta)$ satisfies initial and boundary conditions automatically. It is thus quite data efficient. The recent Physics Informed Neural Network \cite{Raissi2017PhysicsID,Raissi2017PhysicsID2,2019JCoPh.378..686R} uses soft constraints where the violation to initial and boundary conditions are also added to the training objective $L = |\hat{L} f| + \beta_1 |\delta_{BC}| + \beta_2 |\delta_{IC}|$.

Some variational functions should obey physical constraints. 
E.g., in solving the Maxwell equations, the magnetic field represented by the DNN should be divergence free. To take into this constraint, the paper "Linearly constrained neural network" uses DNN to produce a vector field $\vec{A}(x, y, z, \theta)$ whose curl $\nabla \times \vec{A}$ are divergence free \cite{Hendriks2020LinearlyCN}.
It is thus also possible to construct a scalar field $\phi(x, y, z, \theta)$ whose gradients $(\partial_x \phi, \partial_y \phi, \partial_z \phi)$ are curl free. Actually, a general method has been developed to construct neural networks with linear constraints. In solving the many body Schrodinger equations, the many fermion wave function should be anti-symmetric.
FermiNet, PauliNet uses Slater determinant to construct DNNs that are anti-symmetric. \cite{PhysRevResearch.2.033429,Hermann2019DeepneuralnetworkSO}
In DFT \cite{PhysRev.136.B864,PhysRev.140.A1133,condon2019surface} and Molecular dynamics \cite{inbook}, the local chemical environment usually have translational or rotational symmetry that is considered using gauge equivalent neural network \cite{PhysRevLett.127.276402}. 
In Lattice Gauge field theory \cite{Creutz1983UNAS}, Gauge equivariant normalizing flows are employed to sample field configurations \cite{Abbott2022SamplingQF}.

In the present paper, we use a monotonic neural network to represent the cumulative distribution function $\int_{-\infty}^{x} f(x') dx'$,
whose first order derivative is the probability density $f(x) = \psi^*(x)\psi(x)$ that gives the ground state wave function.The present paper demonstrates that neural network with physical constraints can be used as efficient trial wave functions of Schrodinger equations. Auto-diff helps to compute the required derivatives of the trial function with respect to the input variables.
In this way, optimizing the violation of the trial function to PDEs solves PDEs to high accuracy. Compared to previous methods, our calculation does not need to calculate any numerical integrals in the whole calculation and the unitary constraint we impose on the variational wave function increases the data learning efficiency. The improved algorithm greatly reduces the amount of computation required to solve the same Schrodinger equation. These advantages make our method more suitable for dealing with many-body state, which require a huge amount of computation.

\section{m\lowercase{ethods}}
The traditional variational method for quantum mechanics \cite{keeble2020machine,saito2018method} usually use given function with unknown parameters as variational function, e.g., $e^{-\alpha r}$ with $\alpha$ one unknown parameter. Different from the previous Variational Artificial Neural Network (VANN) applications \cite{Hermann2019DeepneuralnetworkSO,carleo2017solving}, we do not use DNN to represent the wave function directly, instead, we use DNN to represent the CDF, which is the integration of the probability density function on the spatial coordinate.The training objective is thus to minimize the violation of the wave function $\psi(x)$ represented by the neural network to the Schrodinger equation.
Its relationship with the wave function are shown as follows,

\begin{align}
F(x) &=\int_{-\infty}^{x} \psi^{\ast } (x')\psi(x')d x' \\
\psi(x)&=\sqrt{\frac{dF(x)}{dx} }  
\label{eq:cdf}
\end{align}
where $F(x)$ is the CDF represented by a neural network that is monotonic by construction.We choose to constrain the weights to make the algorithm data efficient.
The derivative $dF/dx$ is calculated using auto differentiation (auto-diff) \cite{paszke2017automatic}, 
that is provided by the deep learning libraries automatically, in analytical precision.
There are two advantages using CDF. First, the wave function extracted from CDF automatically satisfies the normalization condition. So there is no numerical integral in the whole calculation.
Second, the values of CDF is in between $(0, 1)$, whose range is much smaller than PDF, 
making the neural network much easier to train under the same learning rate. In practice, our training epochs are far fewer than the previous algorithm. And because we eliminate all the integrals, the computation of each epoch is also less than previous algorithms. So our algorithm can achieve higher accuracy with less computation.

We use feed forward neural network, or simply multi-layer percepton to represent the CDF.
The input of the neural network is the n-dimensional spatial coordinates $x$, 
the first layer of the DNN consists $m=32$ hidden neurons whose values are calculated by $h_1 = \sigma(x W_1 + b_1)$,
where $W_1$ is the weight matrix with $n\times m$ elements and $b_1$ is the bias vector with $m$ values.
The $\sigma$ is the activation function which brings neural network non-linear representation ability.
To increase the representation power, the values of neurons in the first hidden layer are feed forward to the 
second hidden layer with similar operations $h_2 = \sigma(h_1 W_2 + b_2)$.
One can stack multiple hidden layers with one output neuron in the last layer to represent the value of CDF function. The whole neural network can thus be thought of as one variational function $F(x, \theta)$ with
$\theta$ all the trainable parameters in the neural network.

To make sure that $F(x, \theta)$ is monotonic, we add a non-negative constraint to the weights $W_i$ of the neural network.
At the same time, the activation function should also be monotonic. 
In principle, sigmoid, tanh as well as leaky relu can all be used to construct this monotonic neural network.
In practice, we use sigmoid activation function whose derivatives are also continuous. 
This is important when the second order derivatives are required in auto-diff.
E.g., if one uses relu activatition function, the second order derivatives of the output of the neural network
to the input equal 0. 
The last layer also uses sigmoid activation function to make sure that the output range is $(0, 1)$.

The training objective is to find the ground state energy $E_0$ and its corresponding wave function $\psi_0$ by minimizing the violation of the wave function $\psi(x)$ to the following Schrodinger equation,

\begin{align}
    H |\psi \rangle = E_0 |\psi \rangle
\end{align}
where $H = -{\hbar^2 \over 2m} \nabla^2 + V({\bf x}) $ is the Hamiltonian operator and $E_0$ represents its smallest eigenvalue. The loss function is thus set to be,
\begin{equation}
L(\theta) =|(H-E_{0})|\psi\rangle +|F(x_{min})|+|F(x_{max})-1|
 \label{eq3}
\end{equation}
where $\theta$ represent all the trainable parameters in the monotonic neural network,
the $\nabla^2 |\psi \rangle$ is computed by the neural network through auto-diff,
$E_0$ is another trainable parameter initialized with constant number $0.0$.
Two additional loss term are added to take into account the boundary condition of the CDF. We use it to limit the range of values of the CDF, which ensures that the wavefunction satisfies the normalization condition.  In previous Variational Artificial Neural Network (VANN) applications, this term was written as $ <\psi|H|\psi>/<\psi|\psi>$. We replace the numerical integration of the denominator with soft constraints, which simplifies the calculation.

Before optimization, the weight values of the neural network parameters are usually initialized randomly  or through Xavier scheme \cite{glorot2010understanding}. In our problem, we observe that the scheme of parameter initialization has little influence on the training process and the result of variational optimization. 

We try to eliminate all numerical integrals in the whole calculation. Because in neural network calculation, the differential is easier to calculate than integral. To calculate the integral, we have to use numerical approximation methods such as Monte Carlo sampling, which will certainly increase the amount of computation and may affect the accuracy. In practice, we found that we had to subtract the energy term if we didn't want to do the integral.To find the ground state energy and wave function, we add another loss term $e^{0.001E_0}$.
The basic logic is to decrease $E_0$ during the optimization of the overall loss.
The function form of this loss term is designed to produce proper negative gradients at different stages of training.
First, the gradients should be large to make it converge fast when $E_0$ is much larger than the ground state energy. Second, the gradients should be small enough to avoid interfering with the optimization of other parts of overall loss when $E_0$ is close enough to the true value. Last but not least, this loss term must monotonously increase with $E_0$ throughout the definition domain. 
In principle, $E_0$ can be smaller than the analytical ground state energy, 
however, in that case the residual of Shrodinger equation increases faster than this term because of the coeficient $0.001$.
After trained, the value of $E_0$ approaches the exact value of the ground state energy. 

We test the performance of the DNN Schrodinger equation solver on three classical quantum mechanical problems.
The first problem is the harmonic oscillator problem \cite{Vasileska2008QuantumMH}. Harmonic oscillator is used to approximate molecular vibration, lattice vibration, radiation field vibration and so on, around the steady point. All of these problems can be regarded as many independent harmonic oscillators whose potential in the Hamiltonian can be written as,
\begin{align}
V=\frac{1}{2}m\omega ^{2}  x^2
\end{align}
where $m$ is the mass of the oscillator, $\omega$ is its angular frequency and $x$ is its deviation from equilibrium position.

The second problem is to solve Schrodinger equation with Woods-Saxon potential \cite{capak2016remarks} that is widely used in nuclear physics to represent the charge distributions of nucleus,
\begin{equation}
V=\frac{-1}{1+e^{\frac{\left | x \right |-R_{0}  }{a_0} } } 
 \label{eq3}
\end{equation}
Where $a_0$ is related to the thickness of the surface layer, in which the potential drops from the outside to the inside of the nucleus and $R_0$ is the average radius of the nucleus at which the average interaction occurs.

The third potential is an infinitely high potential well with a width of $2 l$,

\begin{equation}
V=\left\{
\begin{aligned}
      & \infty,   & \;    & |x| >l \\
      & 0,        & \;    & |x| \le l \\
\end{aligned}
\right.
\label{eq4}
\end{equation}

For the sake of brevity, the parameters in Hamiltonian use the following values, 
\begin{align}
    \hbar = m = \omega = 1, R_0 = 6.2, a_0 =0.1, l = 4.
\end{align}

Different from previous studies that solve Schrodinger equations using supervised learning, our method is close to unsupervised learning where both $E_0$ and $\psi_0$ are learned through optimization.
The input to the neural network is a list of shuffled coordinates sampled from the domain.
Using these coordinates, we compute the loss functions and minimize the violation of $E_0$ and $\psi_0$
to the Schrodinger equation, as well as $e^{0.001E_0}$.
In principle, we can use markov chain monte carlo (MCMC) method \cite{wang2011markov} to sample coordinates with the learned wave function, or use active learning to generate coordinates that violate Schrodinger equations more with the currently learned network, to speed up the training process.
In practice, for these simple problems, the wave function are usually very close to the exact wave function after training the DNN for 2000 iterations. 
We generally train 10,000 iterations with a very small learning rate for the last 1000 iterations.

We use tensorflow \cite{abadi2016tensorflow} to construct the DNN, to compute the auto-diff $dF/dx$ as well as $\nabla^2 \psi$ and to update the network parameters. We use Adam algorithm \cite{kingma2014adam} that add momentum mechanism and adaptive learning rate to the simple stochastic gradient descent $\theta_{n+1} = \theta_n - lr {1\over m} \sum_{i=1}^m {\partial l_i \over \partial \theta}$. The relevant parameters in Adam algorithm are set to $\beta_1=0.9, \beta_2 = 0.999, \epsilon=10^{-7}$.
To speed up the training process, we use learning rate scheduler to adjust the learning rate and make it vary between $10^{-2}-10^{-5}$. A large learning rate at early stage makes the function converge faster at the beginning, and a small learning rate at late time makes the training process smooth. 

To quantify the difference between the true wave function $\psi_{\rm true}$ and the wave function learned by the DNN $\psi_{\rm DNN}$, we introduce partial-wave fidelity K in the following,
\begin{equation}
K=\frac{< \psi_{\rm true}|\psi_{\rm DNN} >^2}{< \psi_{\rm true}|\psi_{\rm true}><\psi_{\rm DNN}|\psi_{\rm DNN} >}
 \label{eq4}
\end{equation}

The closer the K is to one, the closer the result of DNN is to the exact wave function.

\section{r\lowercase{esults} }

For the harmonic oscillator potential used in the present paper, the analytical ground state energy is $0.5\hbar\omega$. 
After 1500 iterations of training, the ground state energy from DNN is $E_0 = 0.50038 \hbar\omega$,
whose relative error is within $0.06\%$.
In the last stage of $10000$ iterations, the error can be controlled below $0.002\%$.

The partial-wave fidelity K is approximately 0.997993 using 3 hidden layers with 32 units (neurons) per layers.
To study the influence of the number of variational parameters on the training results, we computed K using different numbers of hidden layers and numbers of units per layer.
The results are shown below.

\begin{table}[!ht]
    \centering
        \begin{tabular}{|c|l l l l}
    \hline
         \diagbox {$N_{unit}$}{$N_{layer}$}& 1 & 2 & 3 & 4  \\ \hline
        4 & 0.9995717 &  0.9999767 & 0.9999705 & 0.9999618  \\ 
        8 & 0.9999416 &  0.9999797 & 0.9999910 & 0.9999932  \\ 
        16 & 0.9999861 & 0.9999923 &  0.9999936 & {\bf 0.9999967}  \\ 
        32 & 0.9999789 & 0.9999909 & 0.9999896 & 0.9999922 \\ 
        64 & 0.9999744 & 0.9999746 & 0.9999903 & 0.9999941 \\ 
    \end{tabular}
 \caption{The fidelities of VANN result,the first row represents the number of hidden layers, and the first column represents the number of units in each layer}
 \label{tab1}
\end{table}

As shown in Table.\ref{tab1}, the highest fidelity happens using 4 hidden layers with 16 hidden neurons per layer,
with $K=0.9999967$ .
For this simple problem, the DNN achieve a very low error in fidelity even with only two hidden layers and four units per layer.
The performance increases with the the number of variational parameters, to approach the exact wave function. However, the performance of DNN saturate or even decrease if there are too many variational parameters.

\begin{figure}[htp]
    \centering
    \includegraphics[width=8cm]{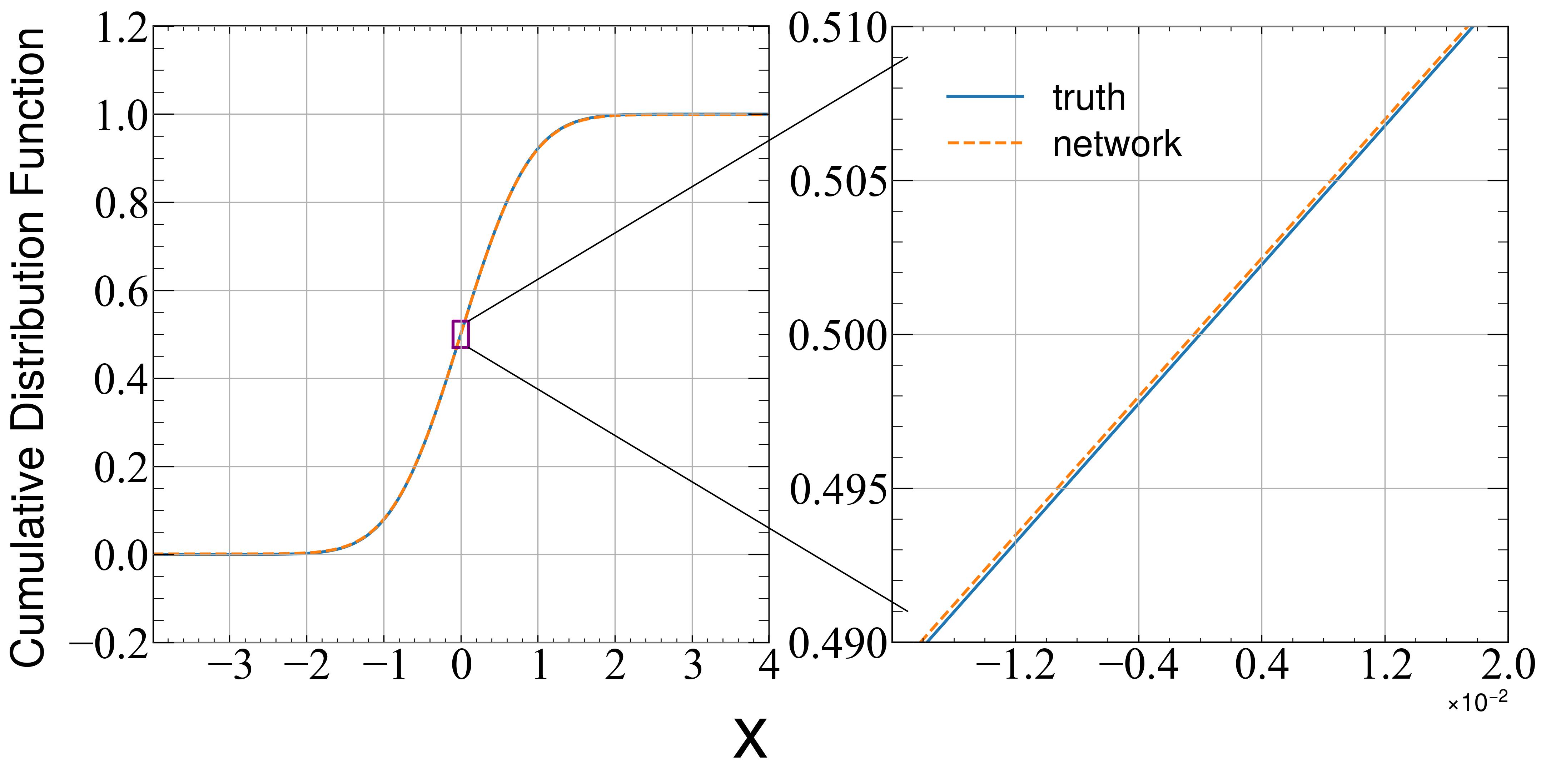}
     \caption{ The cumulative distribution equation as a function of position in harmonic oscillator problem}
     \label{fig:cdf}
\end{figure}

To further visualize the difference between the result of DNN and the exact solution, we compare the CDF in FIG.\ref{fig:cdf}. 

Using Eq.~\ref{eq:cdf}, we compute $\psi(x) = \sqrt{dF/dx}$ and compare the wave function $\psi(x)$, its first and second derivatives ${d\psi \over dx}$ and ${d^2 \psi \over dx^2}$ with the analytical results.
This provides a detailed comparison that shows the power of the variational function represented by the DNN.

\begin{figure}[htp]
    \centering
    \includegraphics[width=8cm]{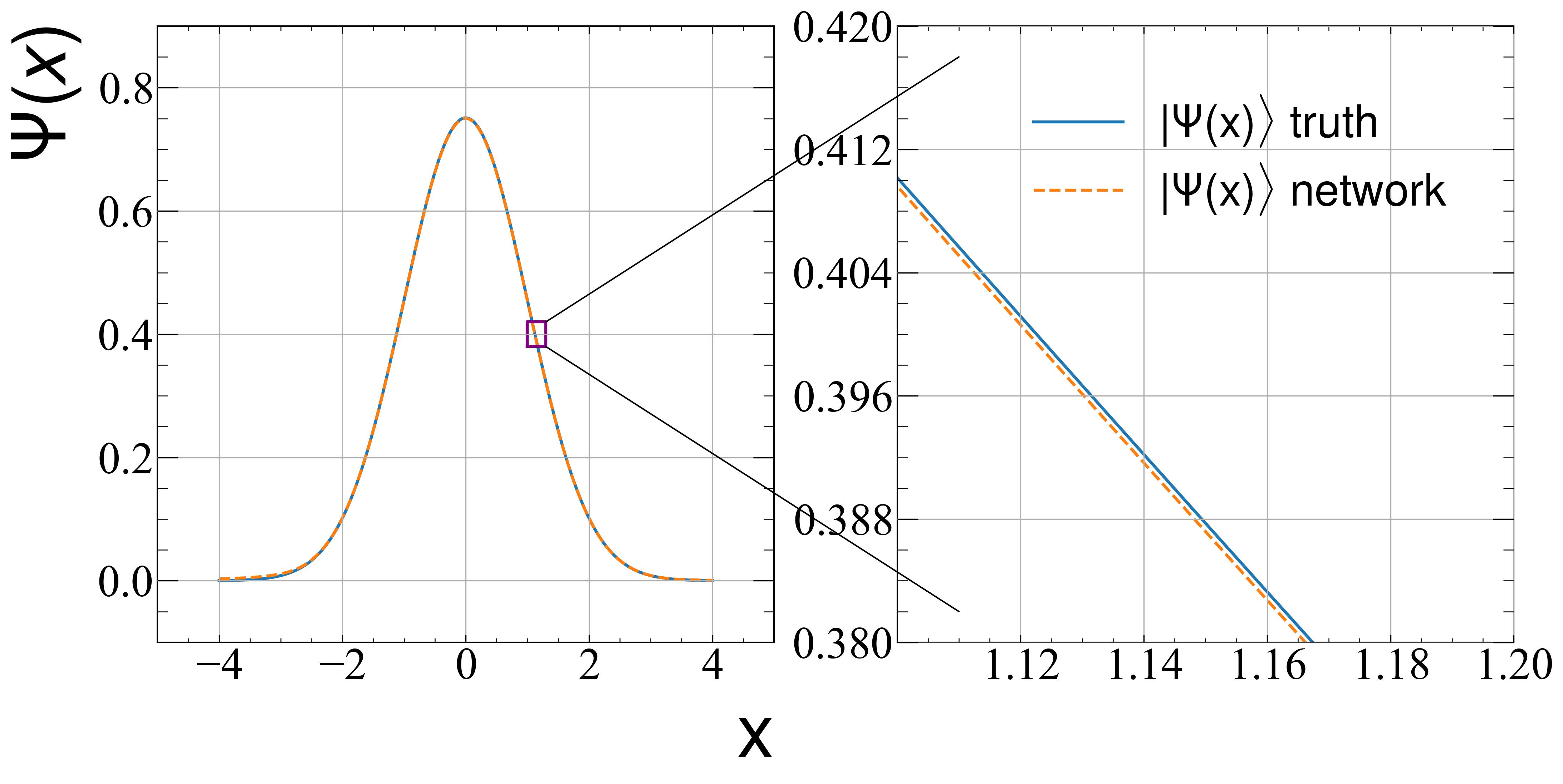}
    \includegraphics[width=8cm]{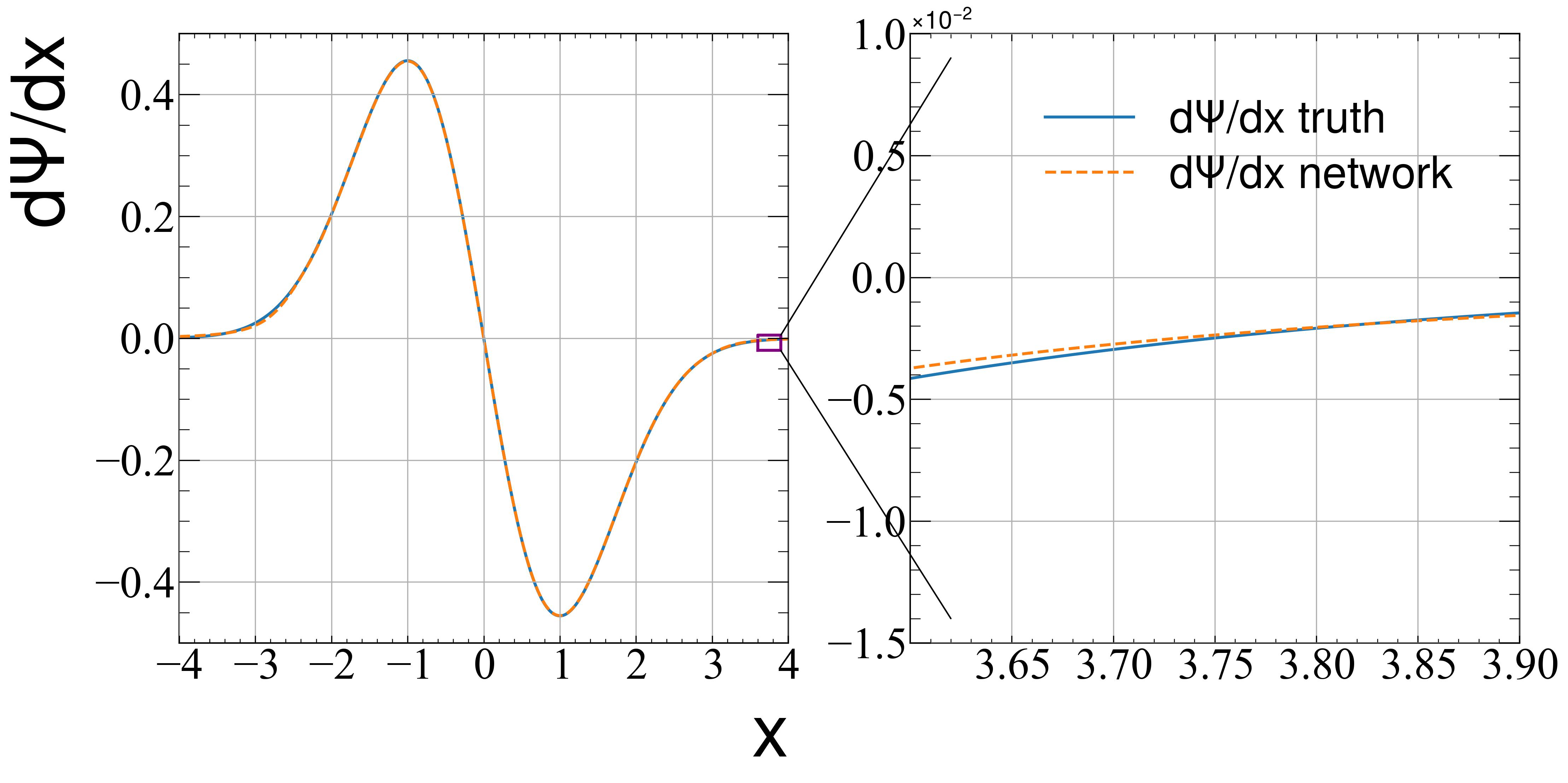}
    \includegraphics[width=8cm]{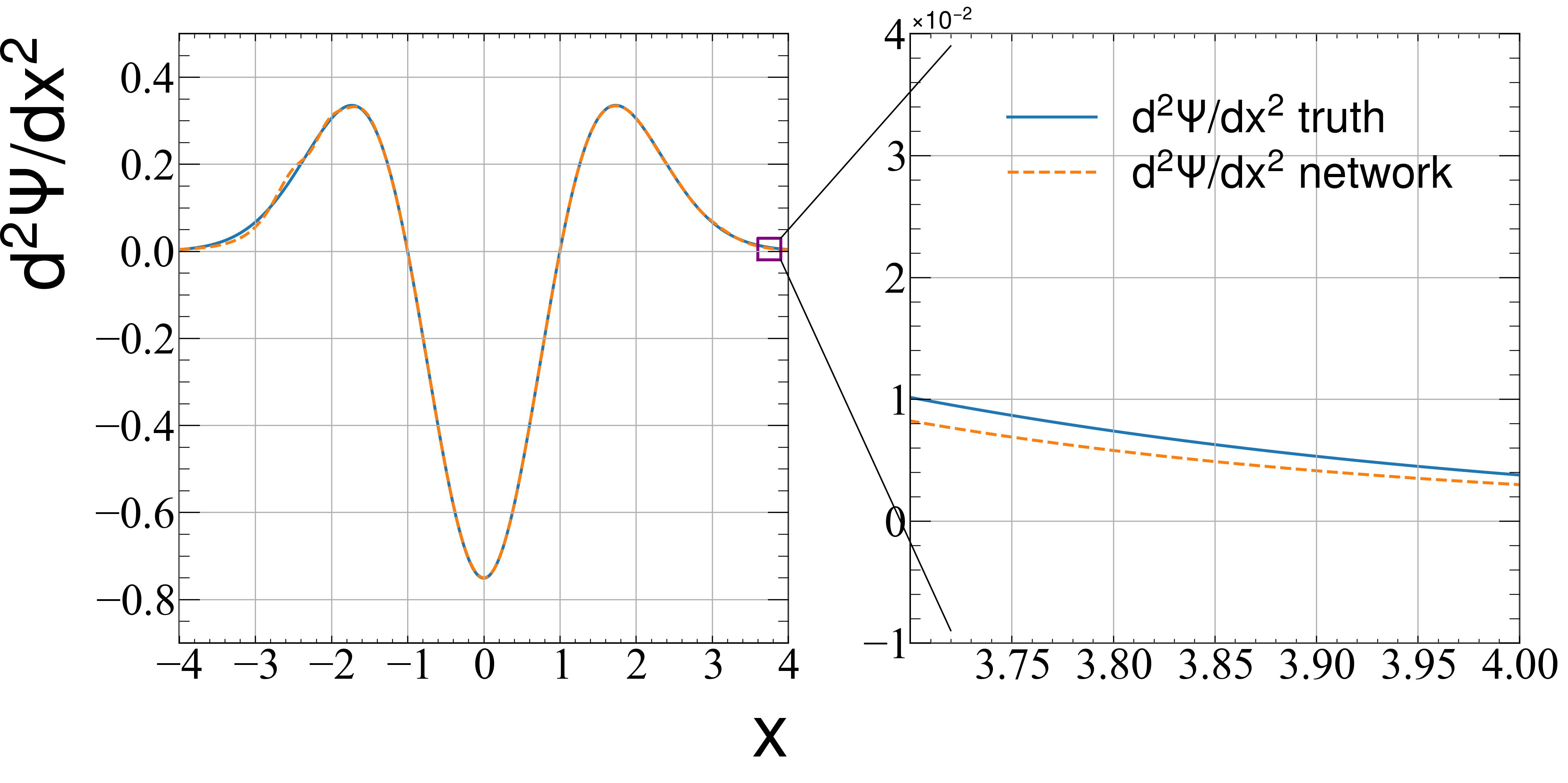}

    \caption{The ground state wave function(top panel),first derivative(central) and the second derivative(bottom) as a function of the position in harmonic oscillator problem. }
    \label{fig:Psi}
\end{figure}

As shown in Fig.~\ref{fig:cdf}, the difference between DNN results and the ground truth is within the error range of 0.0001. The error range of the ground state wave function can be controlled within 0.0002 as shown in Fig.~\ref{fig:Psi}. In addition, the accuracy of the learned first and second order derivatives through variational optimization are also very high, which means that this method is not only accurate, but also captures the true physics in stead of finding an approximation function for the grond state wave function.

\begin{figure}[htp]
    \centering
    \includegraphics[width=8cm]{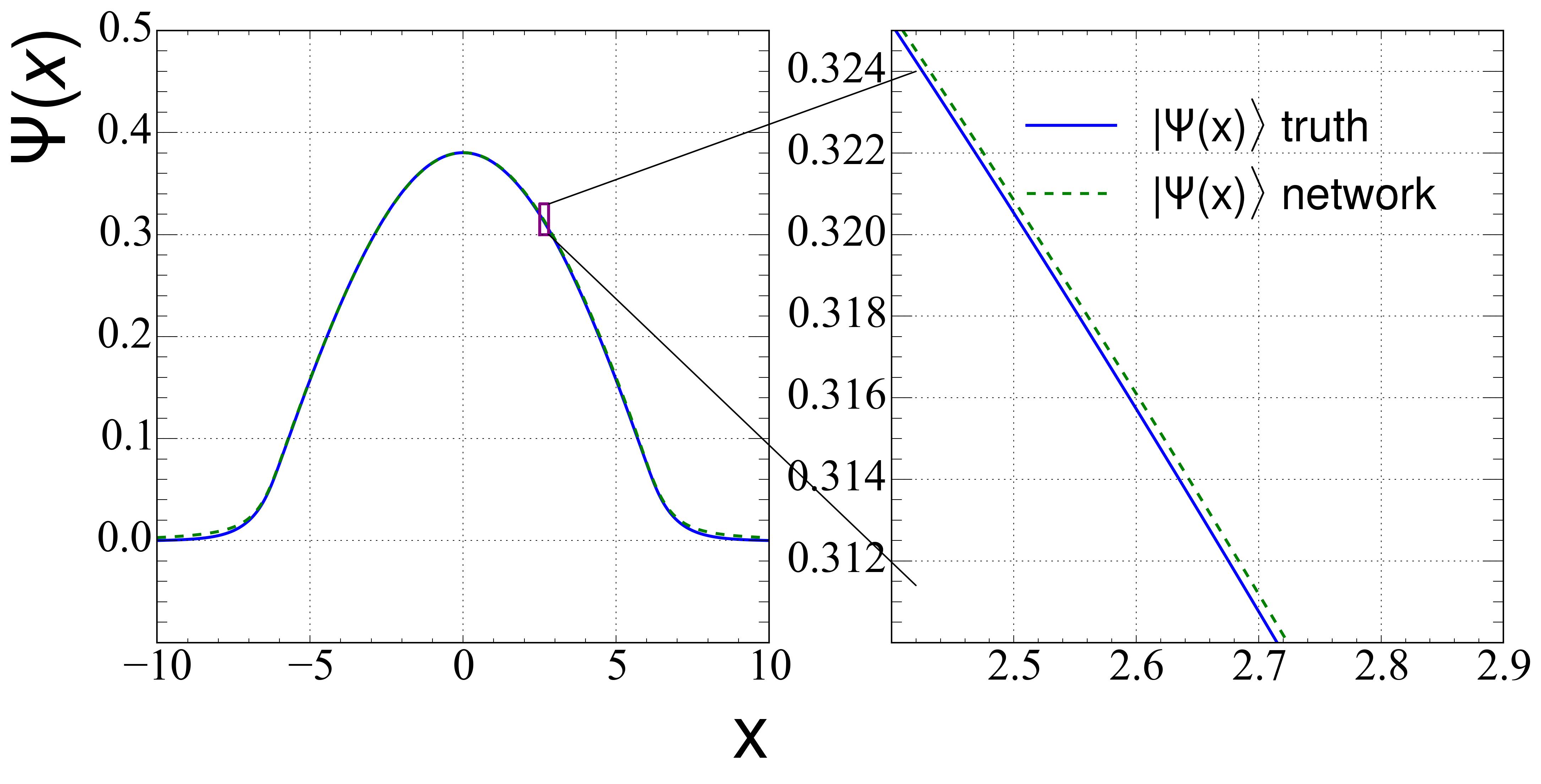}
     \caption{The ground state wave function in the Woods-Saxon potential energy}
     \label{fig:WS}
\end{figure}

\begin{figure}[htp]
    \centering
    \includegraphics[width=8cm]{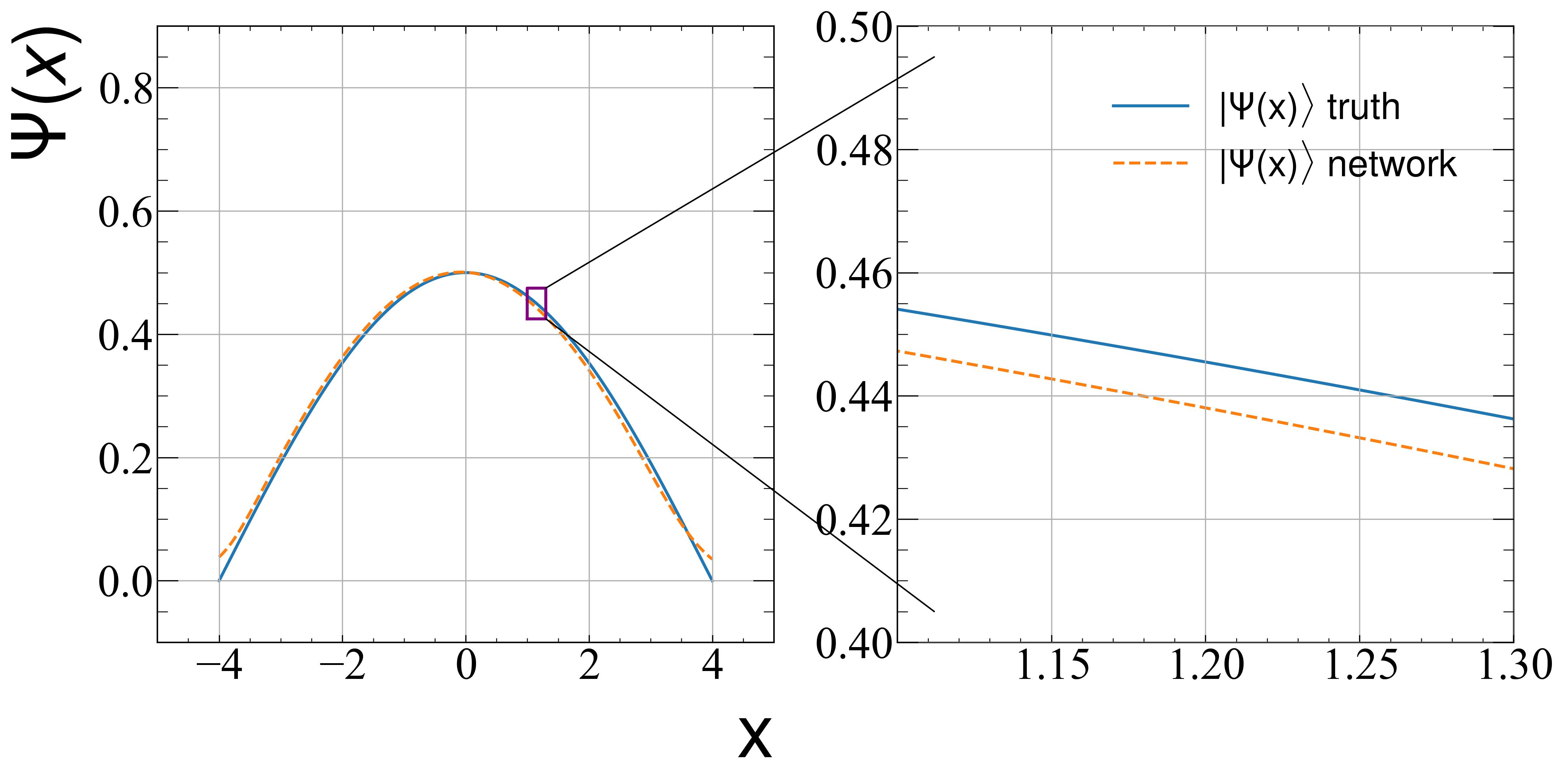}
     \caption{The ground state wave function in infinitely high potential well}
     \label{fig:IW}
 \end{figure}

The same network is used to to solve Schrodinger equations with Woods-Saxon potential and the infinitely high potential well. 
No modification is made to hyperparameters other than the potential  part in the Hamiltonian.
The comparisons between the learned ground state wave functions and the true values are shown in Fig.~\ref{fig:WS} and Fig.~\ref{fig:IW}. 
The result shows that the ground state wave functions obtained from the DNN CDF are also in excellent agreement with the exact solution. The error range of the Woods-Saxon potential's ground state wave function can be controlled within 0.0002. 
As shown in Fig.\ref{fig:IW}, the performance of the network on infinitely high potential well is relatively poor, whose range of error is expanded to 0.02, which is much worse than harmonic oscillator potential and Woods-Saxon potential. This is also reflected in the calculation of ground state energy and the partial-wave fidelity K. The ground state energy calculated by DNN for the Woods-Saxon potential problem is -0.97382, also within an error of $0.002\%$ to the exact result -0.97385 and $K=0.999964$, similar to the harmonic oscillator potential. While for the potential well problem, the DNN $E_0 = 0.07787$, whose relative error to the accurate result 0.07710 is about $1\%$ and $K=0.9977475$ also less than the average level of the first two potentials. It reminds us that this DNN might not perform good at dealing with potentials with discontinuities, like infinitely high potential well, whose potential energy discontinuously change from zero to infinity at the boundary. We think that this is due to the fact that the soft constraint cannot handle the infinite potential energy at the boundary well, so the probability density at the boundary is not 0. 

                                                                                                                                                                                                                                 \section{c\lowercase{onclusions} }

In the present paper, we use physics based neural network to solve Schrodinger equations numerically.
We designed a monotonic neural network to represent the CDF of the ground state wave function.
In this way, the wave function represented by the DNN satisfy the normalization condition by design.
The variational optimization is reduced to a optimization problem by minimizing the violation of the trial wave function and trial ground state energy $E_0$ to Schrodinger equations.
The method is used to solve Schrodinger equations with 3 different potentials, the harmonic oscillator, the Woods-Saxon potential and the infinitely high potential well, all with small relative error.

Compared to traditional variational methods in solving quantum mechanical problems, 
the trial wave function represented by DNN do not have fixed function forms before training.
On the other hand, the training objective is different from the traditional $E_0 = {\langle \psi | H | \psi \rangle \over \langle \psi | \psi \rangle }$ where numerical integration is required for both the numerator and the denominator. 
In our case, the objective is to minimize the violation to the Schrodinger equation, on sampled spatial coordinates.
As the neural network is constrained, the trial wave function is normalized by construction.
Our method is also different from the previous Schrodinger equation solver using supervised learning, 
where ground state energies from numerical solutions are needed to train the neural network.
In another DNN Schrodinger solver \cite{keeble2020machine,saito2018method}, the initial values of the network parameters greatly affect the optimization results. To avoid strong fluctuations, they provide a trial wave function whose form is close to the exact solution. 
The disadvantage of the previous algorithms is that it can only solve the problem that the form of exact solution of the equation is known. Our algorithm can directly ignore the pre-training process, so we do not need to know any information of the exact solution before training, which is more universal and provides the possibility to solve problems that have never been dealt with before.
In addition, we observe that our DNN can approximate the ground state wave function with fewer trainable parameters. 
And the physical constraints constructed in the neural network make the current method quite data efficient. So that we can achieve higher accuracy with less computation.

The current method can be improved in several ways.
First, the CDF works for wave functions in high dimensional space as long as n-dim spatial coordinates are flattened. 
Second, the spatial coordinates used for training can be sampled using the learned wave function or through active learning, to increase the training efficiency.
Third, the anti-symmetric constraints of the wave function should be considered for many fermion systems.
Although further efforts have to be done to improve the current method,
it shows good properties in solving classical quantum mechanical problems.
The next step is to solve the ground state energy and wave functions of Deuteron.
It also paves a new way in solving many nucleon problems.



\section*{Acknowledgments}
This work is supported by the National Natural Science Foundation of China (Grants No. 12035006, No.12075098), the Natural Science Foundation of Hubei Province (Grant No.2019CFB563), the Hubei Province Department of Education (Grant No. D20201108), Hubei Province Department of Science and Technology (Grant No. 2021BLB171). LG Pang and KF Pu also acknowledge the support provided by Huawei Technologies Co., Ltd. Computations are performed at the NSC3 super cluster at CCNU and High-Performance Computing Center of Wuhan University of Science and Technology.
\bibliography{ref}
\end{document}